\documentclass[conference]{IEEEtran}
\IEEEoverridecommandlockouts
\usepackage{booktabs} 
\usepackage{enumitem}
\usepackage{upgreek}
\usepackage{lipsum}
\usepackage{graphicx}
\usepackage[ruled]{algorithm2e} 
\usepackage{cite}
\usepackage{amsmath,amssymb,amsfonts}
\usepackage{algorithmic}
\usepackage[table]{xcolor}
\usepackage{url}
\usepackage{graphicx}
\usepackage{textcomp}

\def\BibTeX{{\rm B\kern-.05em{\sc i\kern-.025em b}\kern-.08em
    T\kern-.1667em\lower.7ex\hbox{E}\kern-.125emX}}

\begin{document}

\title{Developing a K-ary malware using Blockchain}

\author{\IEEEauthorblockN{Joanna Moubarak}
	\IEEEauthorblockA{\textit{ESIB, USJ} \\
		\textit{CIMTI}\\
		Beirut, Lebanon \\
		joanna.moubarak@net.usj.edu.lb}
	\and
	\IEEEauthorblockN{Eric Filiol}
	\IEEEauthorblockA{\textit{ESIEA} \\
		\textit{$(C + V)^O$  Lab}\\
		Laval, France \\
		efiliol@netc.fr}
	\and
	\IEEEauthorblockN{Maroun Chamoun}
	\IEEEauthorblockA{\textit{ESIB, USJ} \\
		\textit{CIMTI}\\
		Beirut, Lebanon \\
		maroun.chamoun@usj.edu.lb}
}

\maketitle

\begin{abstract}
 Cyberattacks are nowadays moving rapidly. They are customized, multi-vector, staged in multiple flows and targeted. Moreover, new hacking playgrounds appeared to reach mobile network, modern architectures and smart cities. For that purpose, malware use different entry points and plug-ins. In addition, they are now deploying several techniques for obfuscation, camouflage and analysis resistance. On the other hand, antiviral protections are positioning innovative approaches exposing malicious indicators and anomalies, revealing assumptions of the limitations of the anti-antiviral mechanisms. Primarily, this paper exposes a state of art in computer virology and then introduces a new concept to create undetectable malware based on the blockchain technology. It summarizes techniques adopted by malicious software to avoid functionalities implemented for viral detection and presents the implementation of new viral techniques that leverage the blockchain network. 
\end{abstract}

\begin{IEEEkeywords}
Malware, K-ary Virus, Malicious program, Blockchain, APT
\end{IEEEkeywords}
\section{Introduction}
Computer infections hit the mainstream in recent years exploiting systems vulnerabilities and creating specific malicious software that are penetrating organizations and governments for damage purposes or to steal information. Also, with the emerging new services and technologies, the marketplace has fully-fledged to reach the cloud, the Internet of Things (IoT) and the interconnected world adding new proliferation environments for malware and augmenting the viral infection risk which is assessed depending on the number of infections and their impacts, the detection ability, the protection in place and the capacity to disinfect and isolate a convicted system.  \newline \indent In 1982, a boot virus was introduced \cite{zett2009nov}. Also, the first official virus appeared in 1983 \cite{zett2009nov} under UNIX proving that no operating system is immune against vulnerabilities. In 1988, viruses and worms leave the laboratories. Since then, malware evolution is exponential. The first viruses and worms for mobile phones appeared in 2004 \cite{chen2004evolution}. In 2008, Stuxnet malware marked a turning point in the enhancement and professionalization of the attacks and the damage they can engender \cite{langner2011stuxnet,farwell2011stuxnet}. Advanced Persistent Threats (APTs) became more omnipresent from 2015, after the hack of Carbanak \cite{abreu2016banking} and Sony Pictures \cite{peterson2014sony} discovered in 2014. In 2016, the Locky ransomware hit considerably several places mainly in Europe \cite{constantin2016new}. Also, the Mirai DDoS attacks on IoT made headlines and derived motives for software developers to incorporate universal security protections \cite{dobbins2016mirai}. Recently, the WannaCry ransomware \cite{greenberg2017wannacry} spreads laterally and a new form of the ransomware called Petya targeted several countries worldwide \cite{richardson2017ransomware}. Furthermore, new Android and iOS malware are discovered each day. These attacks summarize the importance of viral threats and illustrate the evolving viral risk. For instance, the level of risk is potentially high for targeted attacks, probably low to medium for other actors. Furthermore, the main risk today consists in the creation of botnets network utilized to create several types of attacks.     
\newline \indent Besides, antiviral industry is constantly enhancing its capabilities to reduce the gap between the detection and networks containment and always striving to mitigate the breach by combining several analysis mechanisms and machine learning algorithms. However, security solutions can only decrease risks without eliminating them. In reality, the attack has more advantages over the defender. The attacker not only always has the initiative but he innovates constantly. As soon a technique is deemed impossible, the attacker will try to bypass it. Moreover, an attacker will usually seek to hide as long as possible what he has managed to do. Furthermore, the attack always has a lead in time \cite{filiol2007techniques}.

This paper is the result of a prolonged survey on viral techniques adopted by malware as we go through several computer virology studies \cite{cohen1987computer}\cite{hoglund2006rootkits}\cite{russinovich2012windows}. Also, some malware samples were examined in real time against several analysis approaches \cite{moubarak2017comparative} and in the other hand, many antiviral solutions were tested in different attack scenarios and networks. Moreover, several blockchain architecture types \cite{moubarak2017tor} were considered while developing the new malware.       

The remainder of this paper is organized as follows: Section \ref{analysis} develops a summary of viral and antiviral techniques. Section \ref{k-ary} concentrates on k-ary malware, followed by the utilization of the blockchain potential to create a k-ary malware in Section \ref{potential}. We conclude this paper and present our future work in Section \ref{conclusion}. 
\section{Technical Analysis}\label{analysis}
The threat landscape is getting more complicated and businesses remain agile. Many challenges continue to strive security strategies in order to add expanded detection capabilities, find a solution that fits into the architectures and stay within acceptable levels of operational risk. This section gives an overview on viral techniques employed and exposes how antiviral solution providers addresse those challenges. 
\subsection{Computer infections}
Traditionally, the term computer virus is misused to generally refer offensive programs \cite{filiol2006computer}. Currently, malicious softwares are categorized depending on several viral mechanisms. L. Adelman \cite{filiol2006computer} divided malware into two disjoint categories: $simple$ and $Self-reproducing$. Simple malware may be alienated to logic bombs and trojan horse. This category installs itself either in a resident, stealth or persistent mode. Whereas, self-reproducing malware try to overlap all or parts of its malicious code into another program. This class encompasses viruses and worms. Computer infections, whether simple or self-replicating, are installed on a system to compromise the confidentiality, the integrity or the availability of the system. Additionally, the main methods of propagation rely on file sharing, network exchanges, P2P, emails communications and downloading. Mobile computing is another vector of propagation, including direct LAN-WAN and smart phones connections. Recently, malware are surrounding the IoT technology recruiting intelligent devices as zombies to conduct attacks. The recent evolution of infective programs has shown that the scope has surpassed the computer to reach exotic platforms and that the threat becomes global. The mechanisms can certainly vary for one system to another. 

There are several definitions of the concept of computer infection, but none is really complete as recent developments are not taken into account. Attacks by computer infections are all based more or less on social engineering. Another important aspect of the mode of action of the infecting programs is the presence of software vulnerabilities that make the exploitation possible independently of the users. 
\subsection{Antiviral Techniques}
Antiviral Techniques can be alienated to the following:
\begin{enumerate}
	\item Static antiviral techniques: These techniques examine the codes without execution. 
	\begin{enumerate}
		\item Viral signatures: This technique looks for any arrangement of bits and instructions that distinguishes a particular program.
		\item Spectral analysis: This technique consists in examining the code functions and instructions. 
		\item Heuristic analysis: This technique studies the performance and the behavior of a particular program based on policies and guidelines. 
	\end{enumerate}
	\item Dynamic antiviral techniques: These techniques execute the code for analysis.
	\begin{enumerate}
		\item Behavior monitoring: Many mechanisms to monitor the related indicators of compromise (IOCs). 
		\item Code emulation: This technique loads the program into a specific memory zone to mimic the code execution.
	\end{enumerate}
	\item File integrity checking: This technique checks any modification in critical files.
\end{enumerate}

Most efficient antiviral solutions are combining several different antiviral techniques to fight against malware \cite{atdMcAfee,PA,FireEye,Kaspersky}. A layered approach is the typical considered strategy. Most vendor funnel out suspect files as they move through the stack, reducing the number of files requiring sandbox analysis. A mix of signatures, reputation, real-time emulation and heuristics enhance protection and identify advanced malware. Moreover, many security solutions has expanded deployment options with virtual and cloud-based offerings. Antiviral solutions \cite{atdMcAfee} start to stop known threats then adds the next layer of defense using machine learning to detect advanced malware with both statistical analysis and behavioral analysis. Static analysis quickly compares features against those of known threats. If the file cannot be confidently convicted, it will be executed for further behavioral analysis limiting the greyware and blocking suspicious activity. Finally, convictions and IOCs are shared for enhanced protection. 
\subsection{Anti-Antiviral Techniques}
In the other hand, in order to hinder analysis, remain undetected and persist in the network, typically malware use passive and active self-defense mechanisms \cite{filiol2012malicious,filiol2006computer,you2010malware}:

\begin{enumerate}[label=\alph*- ]	
	\item {Stealth techniques}: The ability to deceive any surveillance of the system by reflecting the image of a normal behavior in order to persuade the absence of any infection.
	\item {Polymorphism}: The ability to modify all or part of its own code to prevent any equivalent patterns.
	\item {Code rewriting}: The capacity to modify the code into corresponding functions.
	\item {Applying encryption techniques}: The procedure of masking the code to complicate the cryptanalysis .
	\item {Code armouring}: A number of mechanisms aiming to interrupt, delay or avoid the analysis and make the detection burdensome. 
	\item {Obfuscation}: The fact to store codes in obscured ways to make forensics more challenging. In a $\uptau$ -obfuscation approach, the procedures remains effective for a given time or for a certain trigger.
	\item {Disrupting antiviral solutions}: Many techniques aiming to modify the functioning of antiviral tools and to block specific security queries.
	\item {Packing}: The process of compressing a software outcomes in a new altered sequence of bytes.
	\item {Anti-debugging techniques}: Many techniques aiming to prevent analysts, obtaining context, attaching files and reversing code and able to detect emulation and virtual machines execution.
	\item {Steganography}: The concealment of the viral payload inside another file or image. 
\end{enumerate}

However, these adopted techniques have many limitations and they are obviously apprehended by most antiviral solutions. First, these mechanisms usually require the combination of several techniques. Furthermore, they are difficult to implement and manage. Moreover, some of these techniques may modify the code by adding random instructions or delay the analysis but the final result is the same and the encryption procedures remain unchanged. 
Besides, malware are 100\% of viral information in a single file. Thus, by analysis, we can speculate their operations. There are endless ways to conceive malware. At the moment, designing a truly advanced malware, which will circumvents the known protections, is a difficult and highly competent task. In the rest of this paper, we will present a new approach of malware conception using blockchain and based on the k-ary concept. 
\section{The k-ary malware}\label{k-ary}
This section presents a new category of malware denoted k-ary malware. As an alternative of holding the whole instructions constituting a malicious program in one file, this category encompasses k separate chunks which constitute a partition of the full code. Each of these programs holds only a subdivision of the instructions and reflects a regular uninfected program. 
\subsection{k-ary malware definition}
The K-ary malware was introduced initially in 2007 \cite{filiol2007formalisation} and has been later validated by several  proof-of-concepts (POCs) \cite{filiol2007formalisation,dalla2011hunting}. The formalization of this new type of malware is generalized from Cohen's model using another approach based on vector Boolean functions in order to study softwares interactions. The modalization has proved that simple and polymorphic/metamorphic infections are one way or another correspondent due to the fact that the full information is accessible after the first step of infection. Whereas, the interesting element in k-ary malware consists in the segregation of information. \newline \indent 
Essentially, a k-ary malware is a combined virus where the viral payload is separated and distributed into k different files \cite{filiol2007formalisation}. Each part looks like an innocent executable file and do not generate any indication of compromise (IOC). 
Two main categories of k-ary codes exists \cite{filiol2006computer}:
\begin{enumerate}[label=(\roman*)]
	\item Class I code: The k parts are working sequentially. Three subcategories are to be considered depending on the relation between the several parties. The execution of these k files can be dependent from the others files ($A $ \space $subclass$), no part is denoting the other ($B$ \space $subclass$) or semi-dependent from their execution ($C $ \space $subclass$). 
	\newline 
	\item Class II code: The k parts are working in parallel. Thus, all chucks have to be available and active in the system at the same period.
\end{enumerate}

Furthermore, the k-ary malware are represented in Van Wijngaarden grammars defining the selection of the malware parts\cite{gueguen2010van}:
\begin{equation*}
\alpha R_m \gamma \Leftrightarrow\{\exists\omega \in (\alpha \otimes \gamma ) | \omega\in C(G_m)\}
\end{equation*}
If the result of the selection function $\otimes$ of two files $\alpha$ and $\gamma$ is a part of the code $C$ generated by the malware $m$ then it is a k-ary code.
\subsection{Complexity}
While Cohen \cite{cohen1987computer} and Adleman\cite{adleman1998abstract} analyzed viruses with analogy to Turing machines and recursive functions and a generalized model for malicious behavior have been defined in \cite{zuo2004some} and \cite{bonfante2006abstract}, these studies do not reflect new malware interactions. For instance, the formalization of combined viruses have been well studied in \cite{filiol2007metamorphism}.
Furthermore, it has been demonstrated that the problem of detecting a k-ary malware is NP-complete \cite{filiolmalware}\cite{de2006depth} and that the presence of all codes in memory in Class II codes and Class I (A and C subclasses) constitutes a flaw, except when using a joint rootkit technology. On the other hand, in \cite{gueguen2010van}, the automatic generation of K-ary codes have been detailed, sustaining their detection difficulties and their complexity. Also, in \cite{jacob2010formalization}, k-ary codes were modulated through Join Calculus and have been demonstrated to be undecidable, except for a calculus fragment not inevitably applicable. \newline \indent Therefore, given the NP-completeness of k-ary codes detection, in order to explore the feasibility of a truly undetectable malware, we will use the concept of combined viruses using the blockchain network.    
\subsection{Implementations}
Multiple proof-of concepts confirmed the complexity of combined viruses for OpenOffice in win32 and Linux environment\cite{de2006depth}. \newline \indent Moreover, the different subclasses were validated in serial (4 $\leq$ k $\leq$ 8) \cite{filiol2007techniques} and in parallel (k = 4) \cite{filiol2007techniques}. For instance, each part has been able to regenerate the missing codes under different nomenclatures. \newline \indent  Furthermore, a k-ary virus was implemented in Python \cite{desnos2009implementation} in order to share a secret key utilized to decipher the viral payload. The first use case randomly divided the key between the different parts. However, this method necessitates the availability of all parts in order to retrieve the payload. Another solution consisted in adopting Shamir's Secret Sharing with Neville/Aitken's algorithm to resolve the key and implement the k-ary virus.   
\section{The Blockchain potential in a k-ary malware}\label{potential}
As stated previously, we will develop a k-ary malware utilizing Blockchain. This section gives an overview on blockchain networks and their features and exposes the new k-ary malware implementation and testing.

\begin{figure*}
	\centering
	\includegraphics[width=15cm]{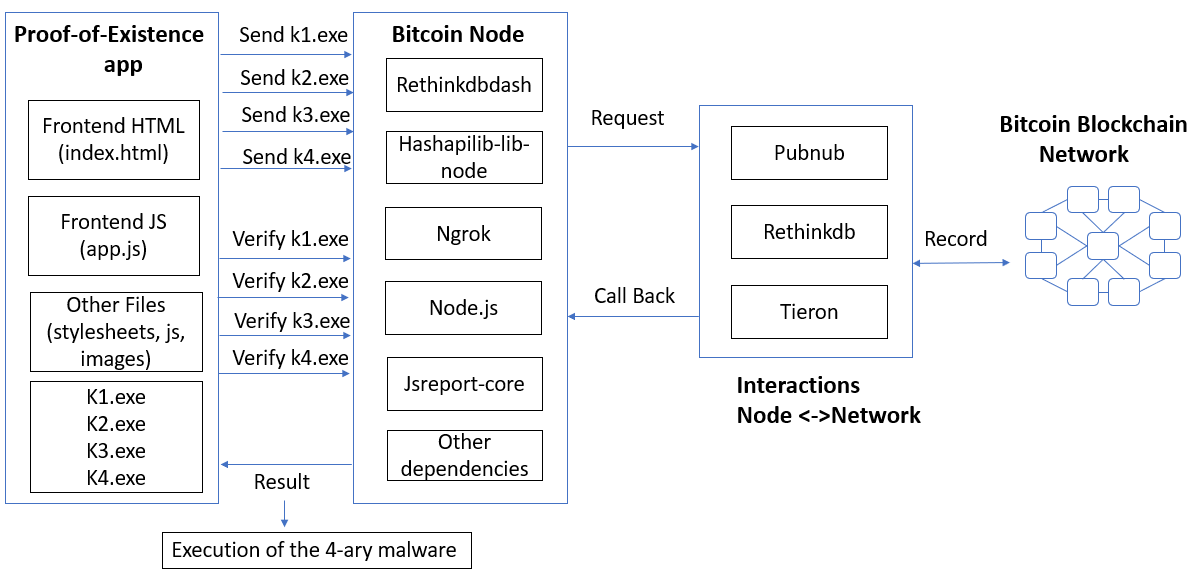}
	\caption{4-ary malware workflow}
	\label{fig:design}
\end{figure*}  
\subsection{Blockchain Overview}
The blockchain is a secure peer-to-peer environment used to maintain a public ledger of transactions between parties where trust is utilized to achieve consensus. This latter depends on several algorithms and typically differ according to the blockchain type and the Distributed Ledger Technology (DLT) employed. Primarly, the blockchain is an immutable data structure using blocks as memory units where each block is referenced by its hash. To characterize the  transactions, the root of the Merkle tree is stored. Each block is composed by several transactions where digital signatures and cryptographic schemes are used to verify each transaction. Moreover, heterogeneous nodes are supported in the distributed network.  Each node will verify and broadcast the block until reaching a consensus. The first miner to validate the blocks is rewarded. Furthermore, many types of consensus algorithms assign a penalty to misbehaving peers \cite{karame2016bitcoin}.  

At the time of writing this paper, there exist more than 700 blockchain types and most of them are alternatives variants of the Bitcoin blockchain. We explored this technology by testing the three mains DLTs in the market nowadays \cite{antonopoulos2014mastering, bahga2017blockchain,IBM} namely Bitcoin, Ethereum and Hyperledger. Mainly, the difference between these networks comes from the fact that Bitcoin and Ethereum are permission-less networks whereas all parties needs to be identified in the Hyperledger blockchain. In addition, the concept of smart contracts, which are function codes compiled with valid transactions, only exists in the Ethereum and Hyperledger networks.
\newline \indent For a long time, the blockchain technology was associated with the Bitcoin DLT based on the Proof-of-Work mechanism. However, each DLT network is caracterized by its own features and concensus algorithm.    

Regardless of the DLT type, the blockchain technology offers numerous security features.

Therefore, the building blocks offered a trusted platform that applications are build on top \cite{bahga2017blockchain,antonopoulos2014mastering}. Some of the blockchain's technology applications are listed bellow:
\begin{itemize}
	\item Proof-of-Existence: Users can verify the existence of a particular content on the blockchain and that it has not been modified. Cryptographic hashes, fingerprint and a proof will be available lastingly.  
	\item Payment Channels: Two parties can exchange transactions ensuring settlement and censorship resistance within a fixed deadline.
	\item Crowdfunding: Users can contribute for many causes and the incremental amount will not be spent until reaching a target.
	\item Event Registration: Users can register to an event or buy tickets through smart contracts interactions. 
	\item IoT: Many IoT applications are taking advantages of blockchain networks in different use cases \cite{IoT}.
	\item Authentication and identification: Many companies are leveraging DLTs for applications and entities validations \cite{LTP}. 
\end{itemize}

The use of blockchain has been beneficial in several applications and in different fields. However, malicious entities took also advantages from this backbone. Cryptocurrencies theft and the 51\% problem where a self-interested miner owns the majority of network work in a Proof-of-Work consensus (in Bitcoin and Ethereum early releases) are typical misuses. Moreover, cryptocurrencies are widely adopted by ransomware infiltrators. And in some cases, malicious contents are sold and uploaded to the blockchain encrypted and abused by the owners of the decryption key further than other mistreated scenarios and dark web applications. Furthermore, the Tor network recently leveraged the blockchain to conduct illegitimate activities \cite{tor}. Besides, in the next section, we will leverage the blockchain as a main entity to create the k-ary malware.

\begin{figure}
	\centering
	\includegraphics[width=\linewidth]{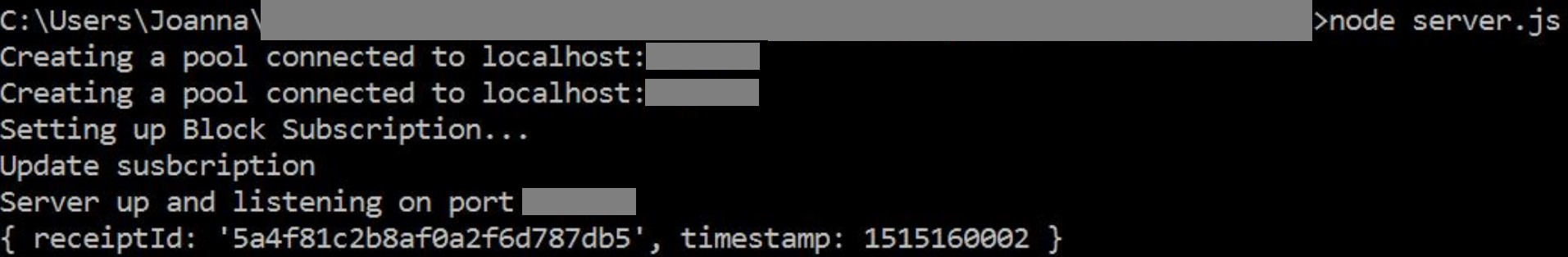}
	\caption{Node.js}
	\label{fig:nodejs}
\end{figure}     
\subsection{Malware design}
As we have seen previously, several attacks scenarios leveredged the blockchain to conduct fraudulent activities \cite{blockchain-ware}. Whereas, for the conception of the new viral algorithms, the blockchain network is a crucial part of the new k-ary malware design. 

Designing a k-ary malware \cite{desnos2009implementation} will lead us to a key management problem to identify each node and a key generation problem to agree on the complexity of the keys. Besides, we need to add randomization for more efficiency. To resolve these problems, we resorted to the blockchain technology. 
\newline \indent In 1988, the authors of \cite{riordan1998environmental} proposed viruses as a solution for handling cryptographic keys. Besides, k-ary viruses are considered for this use case as well where the encrypted payload is confined in one part and the secret key available in another part \cite{filiol2006computer}. As for the proposed new viral algorithms, we have leveraged the cryptographic schemes, the hashing functions, and the digital signatures, which characterize the blockchain network, to develop the new k-ary malware.

\begin{figure}
	\centering
	\includegraphics[width=\linewidth]{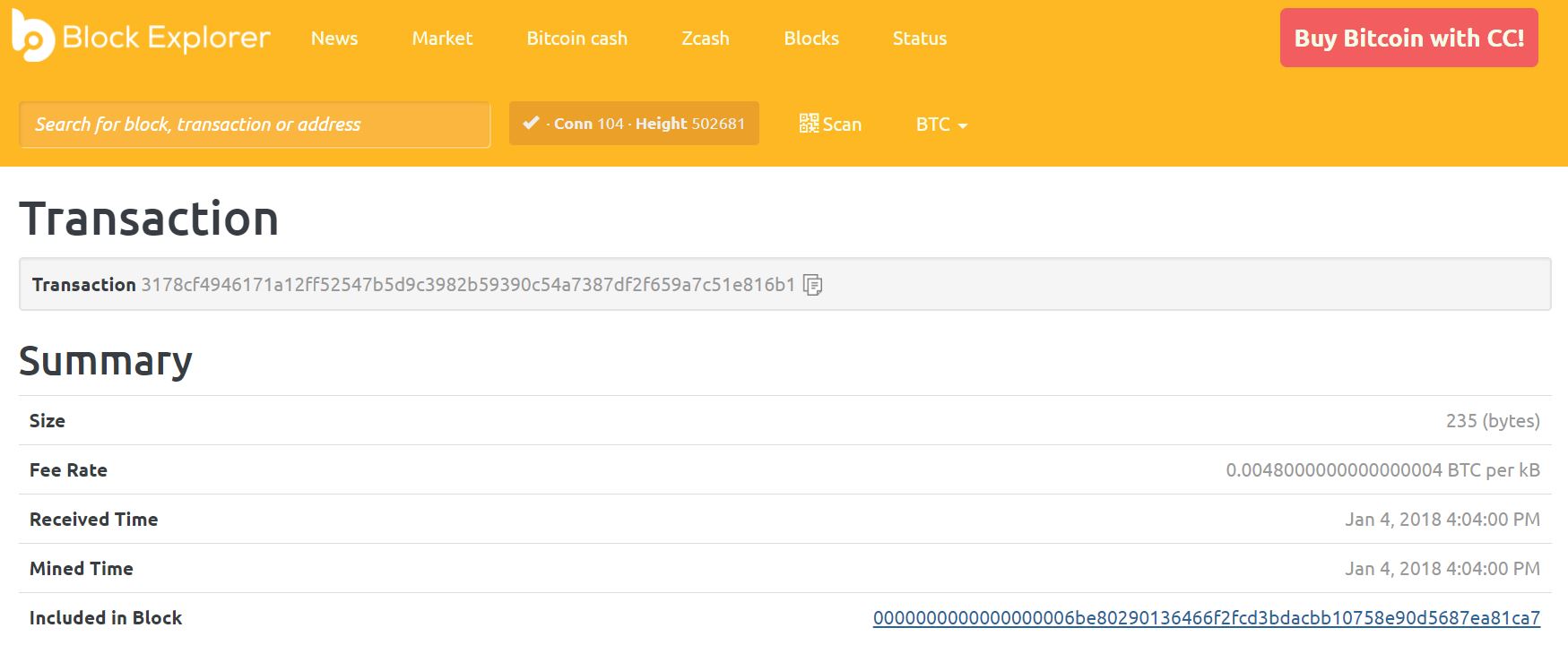}
	\caption{k1.exe transaction summary.}
	\label{fig:explorer0}
\end{figure}

\begin{figure}
	\centering
	\includegraphics[width=\linewidth]{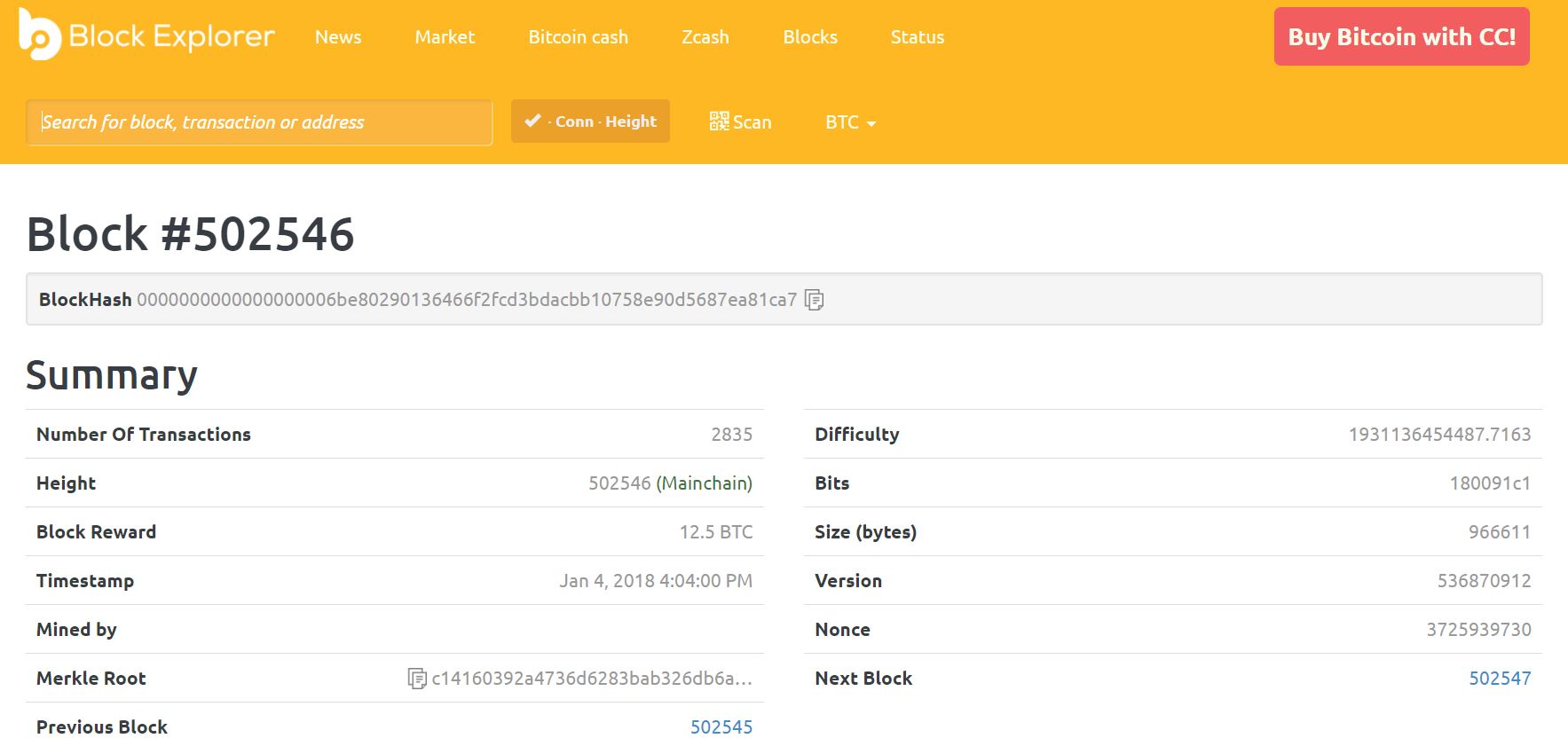}
	\caption{k1.exe block summary.}
	\label{fig:explorer1}
\end{figure}
\begin{figure}
	\centering
	\includegraphics[width=\linewidth]{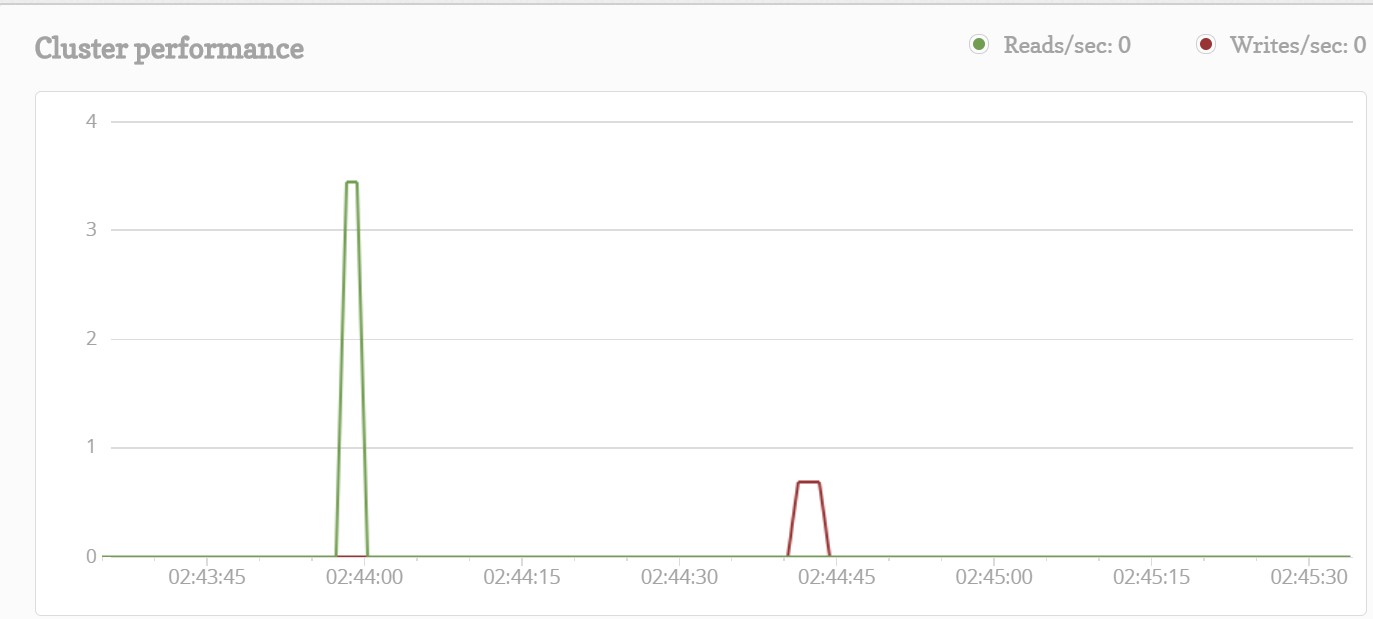}
	\caption{RethinkDB activity.}
	\label{fig:rethinkdb}
\end{figure}
\begin{figure}
	\centering
	\includegraphics[width=\linewidth]{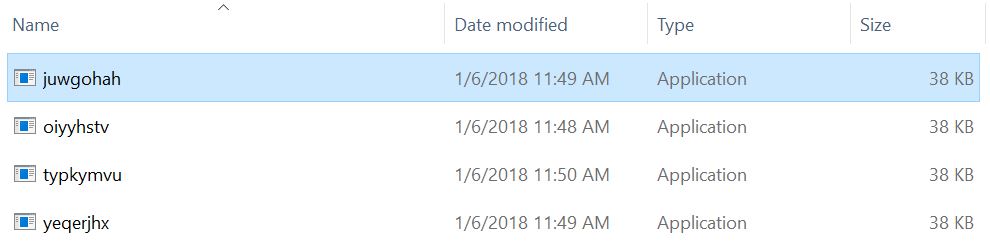}
	\caption{k1 Execution: Auto-reproduction}
	\label{fig:autoreproduction}
\end{figure}

The key components of the malware include a proof of existence application (see Fig. \ref{fig:nodejs}) that interact with the Bitcoin Blockchain Network through the integration of Pubnub, Rethinkdb and Tieron platforms. A detailed tutorial for this combination is given in \cite{poe} to which we referred to in the implementation. Fig. \ref{fig:design} shows the malware workflow.
\newline \indent For the new k-ary algorithmic, the viral payload is splitted in 4 different files. The first viral mechanism employed is the auto-reproduction property generated by k1.exe. The second k-ary file include a keylogging action. The third executable file encompass the property to hide a specific file and the final k-ary executable permits the auto-execution at system startup. Alternatives or additional malicious activities can also be used. For example, the need to interact with a command-and-control server may also be written in a segregated file and added to the design. Also, the auto-deletion propriety implementation provides an interesting feature. Furthermore, breaking up more the code can add more stealthiness. For this POC, a 4-ary malware was tested. The next step consists in submitting the content to the blockchain which store a record of each file that anyone can verify its existence at any time in the blockchain explorer (see Fig. \ref{fig:explorer0} and Fig. \ref{fig:explorer1}). Moreover, the hashing of each executable and the receipt that were given by the blockchain, will be recorded in Rethinkdb \footnote{An open-source database with real time capabilities.} (see Fig. \ref{fig:rethinkdb}) through Tieron\footnote{A helper platform to manage blockchain requests. } APIs and Pubnub\footnote{A data stream network.} real-time processing. Furthermore, the ngrok service provides a secure tunnel to connect with Tieron and receives callbacks. 
\newline \indent In order to execute the 4-ary malware, we verify the existence of each file in the bitcoin network through the hashing signatures and if confirmed, we execute the viral payload (see Fig. \ref{fig:autoreproduction}). In the testing scenario, we created a class I independent (B subclass) 4-ary malware which is the most complex class in term of detection because no executable helps to spot the other \cite{filiol2006computer}.

\subsection{Attack}
Our proof-of-concept is based mainly on the proof-of-existence application. Therefore, in order to convict other systems, a medium to interact with our malicious application is needed. Phishing or other techniques can be employed for that purpose. Besides, the core application databases and flows subscriptions are scheduled for some period of time and the accounts utilized are removed. This makes the attacker anonymous.    
\section{Conclusion}\label{conclusion}
According to antiviral solutions editors, they are able to detect 99.9999 per cent of known and unknown malware. However, in 1986, F. Cohen proved that the detection of viruses is an undecidable problem \cite{filiol2006computer}. \newline \indent Many defense strategies are deployed nowadays to reduce the level of accepted risk. Primary, network traffic analysis is used to create a baseline for ordinary network flows and spot anomalies. Also, full-packet capture mechanisms are deployed for better visibility, reporting and network forensics. Furthermore, payload and behavioral analysis in sandboxes are considered for advanced malware discovery. Moreover, application containment approaches are employed through agents to exclude potential offensive programs in containers and intercept their malicious activity. Finally, many agents are used for data collection and endpoints monitoring using intelligence to provide efficient protection and incident response.
To fight against computer infections, several combination approaches are essentials to eliminate potential intrusions. Also, solutions integration is crucial for management, insight, activities linking and security coverage. Although these solutions are essentials, their scope of operation is limited. Furthermore, the usefulness of the behavioral detection and the will to obtain a lower probability as well as the compromises and algorithmic choices may completely inhibit the essential property of the detection. The fundamental lever is the human factor and the security policies in place. According to the theory of computability, some problems are not calculable and the problem of viral detection is one of the undecidable problems.\newline \indent 
In this paper, we utilized the Blockchain in order to explore the feasiblity of a new undetectable malware.  We based our malware on k-ary codes which have been demonstrated to be NP-complete. We have developed a 4-ary malware and tested it in real time where each chunk of the code interacts with the Bitcoin network to be validated and to make sure that it belongs to our malicious software. Therefore, the blockchain network provided an elegant solution to retrieve the multiple parts of the malware, making sure of their authenticity and integrity without worrying about the generation, the management and the storage of the keys. \newline \indent The next step consists in leveraging smart contracts functions to enhance our k-ary malware and add more complexity. Besides, we will tackle its formalization and validate it in real time against advanced and sophisticated endpoints protections.  
\bibliography{Library1}

\end{document}